\documentclass[]{spie}  
\usepackage[]{graphicx}
\usepackage[]{subfigure}
\usepackage[]{lscape}

\newcommand{\bicep}{B{\sc icep}}

\newcommand{\dasi}{DASI}
\newcommand{\spider}{S{\sc pider}}
\newcommand{\keck}{Keck Array}

\newcommand{\etal}{\emph{et al.}}
\newcommand{\cryomech}{Cryomech}

\newcommand{\lsim}{\lower 2pt \hbox{$\, \buildrel {\scriptstyle <}\over{\scriptstyle \sim}\,$}}

\title{BICEP2 and Keck Array operational overview and status of observations} 

\author{R. W. Ogburn IV\supit{$\dagger$a,b},
        P. A. R. Ade\supit{c},
        R. W. Aikin\supit{d},
        M. Amiri\supit{e},
        S. J. Benton\supit{f},
        C. A. Bischoff\supit{g},\\
        J. J. Bock\supit{h,d},
        J. A. Bonetti\supit{h},
        J. A. Brevik\supit{d},
        E. Bullock\supit{i},
        B. Burger\supit{e},
        G. Davis\supit{e}\\
        C. D. Dowell\supit{h,d},
        L. Duband\supit{j},
        J. P. Filippini\supit{d},
        S. Fliescher\supit{i},
        S. R. Golwala\supit{d},
        M. S. Gordon\supit{g},\\
        M. Halpern\supit{e},
        M. Hasselfield\supit{e},
        G. Hilton\supit{k},
        V. V. Hristov\supit{d},
        H. Hui\supit{d},
        K. Irwin\supit{k},
        J. P. Kaufman\supit{l},\\
        B. G. Keating\supit{l},
        S. A. Kernasovskiy\supit{a},
        J. M. Kovac\supit{g},
        C. L. Kuo\supit{a,b},
        E. M. Leitch\supit{m},
        M. Lueker\supit{h,d},\\
        T. Montroy\supit{n},
        C. B. Netterfield\supit{f},
        H. T. Nguyen\supit{h,d},
        R. O'Brient\supit{h,d},
        A. Orlando\supit{l},
        C. L. Pryke\supit{i},\\
        C. Reintsema\supit{k},
        S. Richter\supit{g},
        J. E. Ruhl\supit{n},
        M. C. Runyan\supit{d},
        R. Schwarz\supit{i},
        C. D. Sheehy\supit{i},\\
        Z. K. Staniszewski\supit{h,d},
        R. V. Sudiwala\supit{c},
        G. P. Teply\supit{d},
        K. Thompson\supit{a},
        J. E. Tolan\supit{a,b},\\
        A. D. Turner\supit{h},
        A. G. Vieregg\supit{g},
        D. V. Wiebe\supit{e},
        P. Wilson\supit{h}, and
        C. L. Wong\supit{g}
\skiplinehalf
\supit{a}Stanford University, Stanford, 382 Via Pueblo Mall, CA 94305, USA \\
\supit{b}Kavli Institute for Particle Astrophysics and Cosmology (KIPAC), Sand Hill Road 2575, Menlo Park, CA 94025, USA \\
\supit{c}Dept. of Physics and Astronomy, Cardiff University, The Parade, Cardiff, CF24 3AA, UK \\
\supit{d}California Institute of Technology, 1200 E. California Blvd., Pasadena, CA 91125 USA \\
\supit{e}Department of Physics \& Astronomy, University of British Columbia, 6224 Agricultural Road, Vancouver, BC V6T1Z1, Canada \\
\supit{f}Department of Physics, University of Toronto, Toronto, ON M5S 1A7, Canada  \\
\supit{g}Harvard-Smithsonian Center for Astrophysics, 60 Garden Street, Cambridge, MA 02138 \\
\supit{h}Jet Propulsion Laboratory, 4800 Oak Grove Dr., Pasadena, CA 91109, USA \\
\supit{i}University of Minnesota, Minneapolis, MN 55455, USA \\
\supit{j}Service des Basses Temp\'eratures, DRFMC, CEA-Grenoble, 17 rue des Martyrs, 38054 Grenoble Cedex 9, France \\
\supit{k}NIST Quantum Devices Group, 325 Broadway, Boulder, CO 80305, USA \\
\supit{l}University of California, San Diego, La Jolla, CA 92093, USA \\
\supit{m}Kavli Institute for Cosmological Physics, University of Chicago, 5640 South Ellis Avenue, Chicago, IL 60637, USA  \\
\supit{n}Physics Department, Case Western Reserve University, Cleveland, OH 44106, USA \\
}

\authorinfo{\supit{$\dagger$}Corresponding author: R. W. Ogburn, 382 Via Pueblo Mall, Stanford, CA 94305. E-mail: ogburn@stanford.edu}

\begin{document} 
  \maketitle 

\begin{abstract}
The \bicep2 and \keck\ experiments are designed to measure the
polarization of the cosmic microwave background (CMB) on angular scales of 2-4 degrees ($\ell=50$--100).  This is the
region in which the $B$-mode signal, a signature prediction of cosmic inflation, is expected to peak.
\bicep2 was deployed to the South Pole at the end of 2009 and is in the middle of its third year of observing with
500 polarization-sensitive detectors at $150~\mathrm{GHz}$.  The \keck\ was deployed to the South Pole
at the end of 2010, initially with three receivers---each similar to \bicep2.  An additional two
receivers have been added during the 2011-12 summer.  We give an overview of the two experiments,
report on substantial gains in the
sensitivity of the two experiments after post-deployment optimization, and show preliminary maps of
CMB polarization from \bicep2.
\end{abstract}



\section{INTRODUCTION}
\label{sec:intro}  
Over the past 15 years, the $\Lambda$CDM theory has emerged as the ``standard model'' of cosmology.
It describes a universe in which cold dark matter drives the formation of structure,
and dark energy (or a cosmological constant) causes accelerating expansion of space.
This concordance model explains the observed universe
well, but leaves several important questions unanswered, such as the nature of the dark energy and dark matter.
It accommodates, but does not predict, the observed flatness, homogeneity, and isotropy of the universe.
These three properties can be explained by the addition to the model of cosmic inflation.  If the universe
went through an inflationary phase in the first instant after the Big Bang, this rapid, exponential expansion
from a single microscopic volume would
have smoothed out spatial curvature.  The observed homogeneity and isotropy arise from the fact that the entire
observable universe would have been in causal contact before inflation.  Quantum fluctuations in the inflaton field
give rise to scale-invariant density fluctuations that would seed structure formation and the primary temperature
anisotropy of the cosmic microwave background (CMB).  
Current CMB observations are in good agreement with inflationary predictions.
\cite{boomerang00,wmapinfl03}

In addition to these \emph{scalar} density perturbations, inflation also generates primordial gravitational waves,
i.e. \emph{tensor} metric perturbations.  Both types of perturbation give rise to temperature variations at the
time when CMB photons last scattered, but with different spatial structure; they accordingly lead to different
signatures in the CMB.  They can be distinguished by decomposing the patterns of CMB polarization into two classes:
the $E$-modes, which have even parity and arise from scalar and tensor perturbations; and the $B$-modes,
which have odd parity and arise from tensor perturbations only.
The detection of $B$-modes would confirm a key prediction of inflation, and a measurement of (or upper limit on)
the amplitude of $B$-modes would provide information about the energy scale of inflation, which determines the
tensor-to-scalar ratio $r$.\cite{primer97}

   \begin{figure}
   \begin{center}
   \begin{tabular}{c}
   \includegraphics[height=8cm]{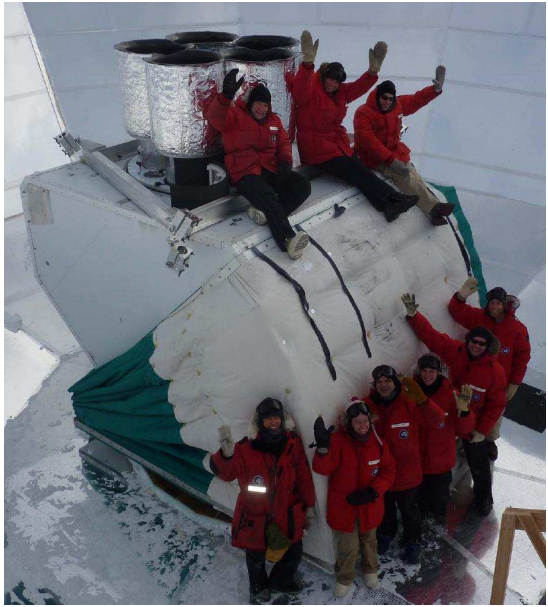}
   \end{tabular}
   \end{center}
   \caption[example]
   { \label{fig:spud5}
\keck\ telescope in February 2012, with five receivers installed in the mount originally built for \dasi\ at the South Pole.
}
   \end{figure}

The \dasi\ experiment measured degree-scale $E$-modes in 2002, the first detection of CMB polarization.\cite{dasi02}
The amplitude of $B$-modes is expected to be lower, and their detection remains an open challenge.  The \bicep2, and
\keck\ experiments have been designed to meet this challenge.
The two experiments adopt elements from the successful \bicep\ telescope, which observed from 2006--08.  \bicep\
has set the most sensitive limits on $B$-mode polarization of the CMB in the range
$\ell=21-335$, and constrained $r<0.72$ at 95\% confidence level.\cite{chiang10,takahashi10}
\bicep2 and the \keck\ use a new detector technology, the Caltech-JPL antenna-coupled TES arrays.\cite{orlando09,orlando10}
These detectors are read out using NIST SQUID amplifiers with the
University of British Columbia MCE control and readout electronics.\cite{mce08}
The monolithic fabrication and multiplexed readout allow \bicep2 and the \keck\ to field larger numbers of
detectors, for improved sensitivity over their predecessors.
\bicep2 was deployed to the South Pole in late 2009\cite{ogburn10,aikin10,brevik10}
and has observed through 2010 and 2011, with a third year of observations now in progress.

With its compact size and low cost, the \bicep-style refracting telescope can be scaled up by
building an array of multiple receivers.  This is the idea behind the \keck\, which uses the
same detector technology as \bicep2 and a similar telescope design, 
with a few additional improvements.\cite{sheehy10}
Instead of consumable liquid helium, the \keck\ uses \cryomech\ pulse-tube coolers to provide cooling at $4~\mathrm{K}$
and above.
It uses the larger mount originally built for the \dasi\ experiment and
attached to the Martin A. Pomerantz Observatory (MAPO).  This mount can accommodate up to five \bicep2-style receivers.
The \keck\ was deployed and commissioned in December 2010--January 2011, with three cryostats housing three full
focal planes of detectors.  It observed with three receivers through 2011, and was upgraded to five receivers for the
2012 season (Fig.~\ref{fig:spud5}).

Both experiments observe the CMB in the same primary field as \bicep1, dubbed the "Southern Hole," with secondary observations of
a bright region of the Galactic plane.

In this paper we present an overview of the \bicep2 and \keck\ experimental design;
a number of improvements to the sensitivity of each experiment that have been made since deployment;
and a look at $E$- and $B$-mode maps from two seasons of \bicep2 observation.
Several companion papers
in this volume present 
the sensitivity of the \keck\ (Kernasovskiy \etal\cite{kernasovskiy12}),
studies of beam shape and differential pointing (Vieregg \etal\cite{vieregg12}),
and further developments in the detector fabrication for the \keck\ and other experiments (O'Brient \etal\cite{obrient12}).

\section{EXPERIMENTAL OVERVIEW}

   \begin{figure}
   \begin{center}
   \subfigure[]{
     \raisebox{3mm}{\includegraphics[height=5cm]{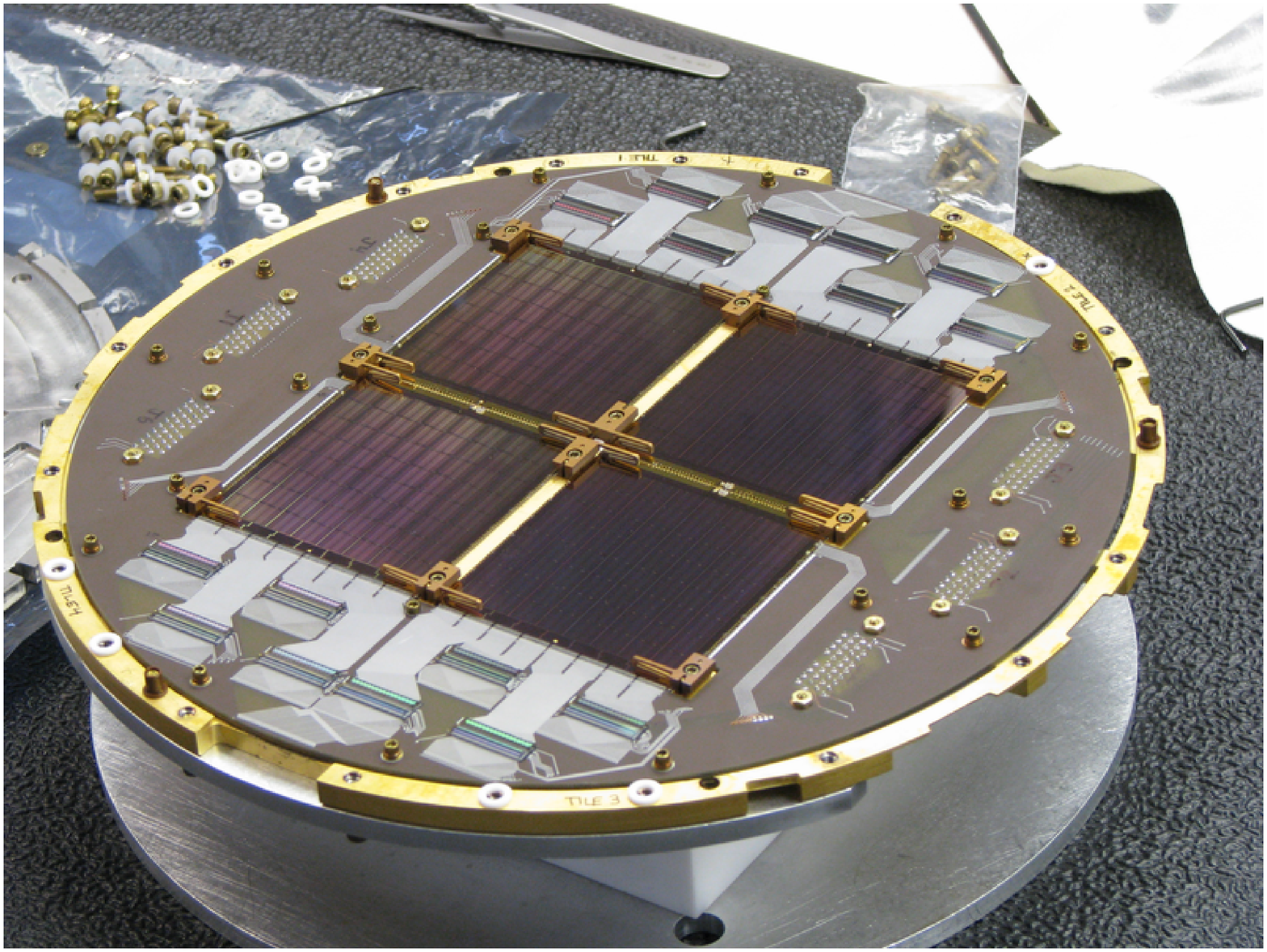}}
     \label{fig:b2fpu}
   }
   \hspace{5mm}
   \subfigure[]{
     \includegraphics[height=5.6cm]{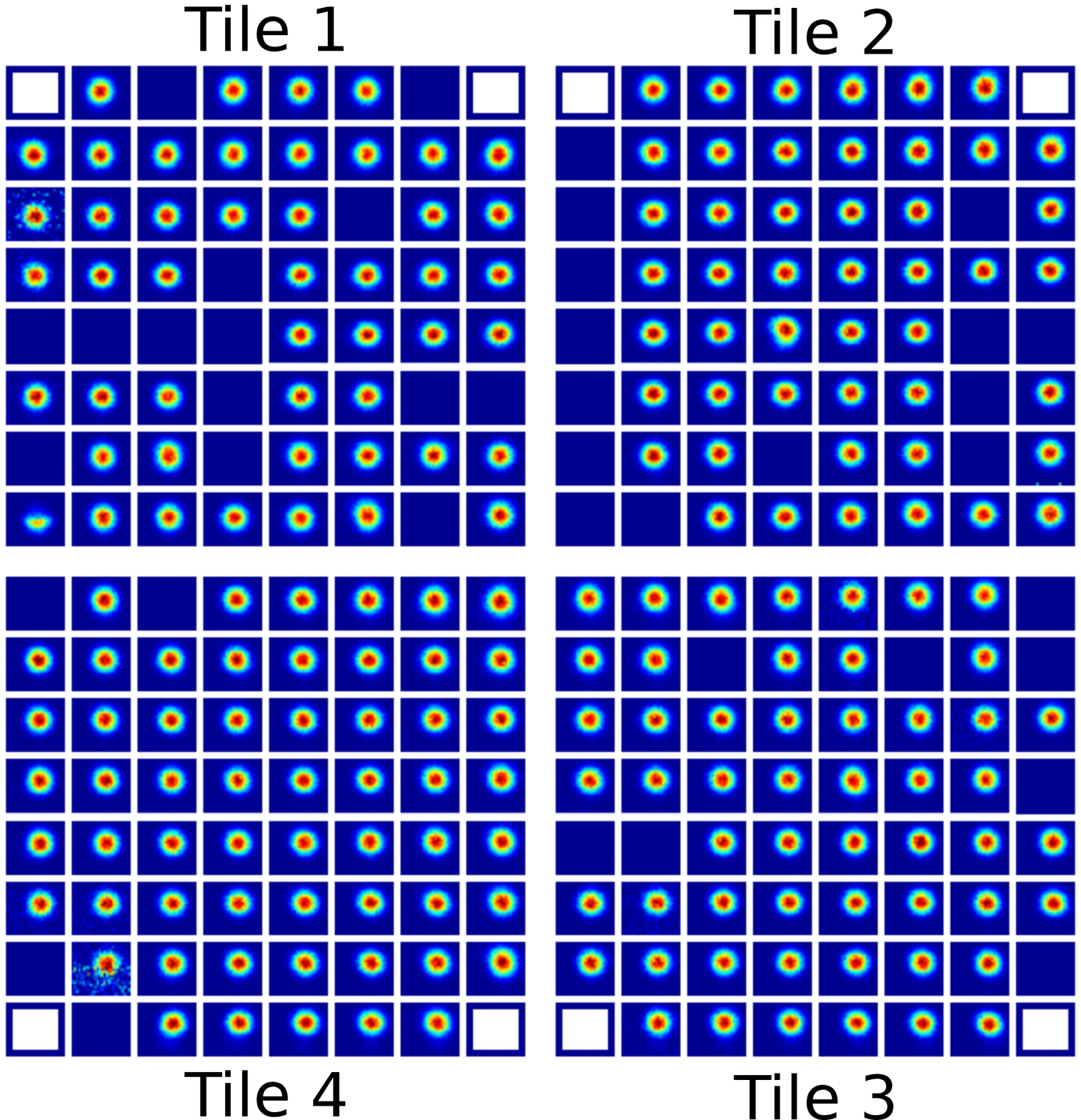}

     \label{fig:b2beams}
   }
   \caption{(a) \bicep2 focal plane with four detector tiles, at the South Pole before installation in
     November 2009.  (b) Measured far-field beam maps for a single \keck\ receiver (rx1).  This image was
     made from data taken at the South Pole in February 2012 using a thermal source raised on a mast.
     Hollow squares represent the location of dark detectors for which the antenna networks were 
     deliberately left disconnected from the TES islands.\label{fig:b2fpubeams}}
   \end{center}
   \end{figure}

   \begin{figure}
   \begin{center}
   \begin{tabular}{c}
   \includegraphics[height=9cm]{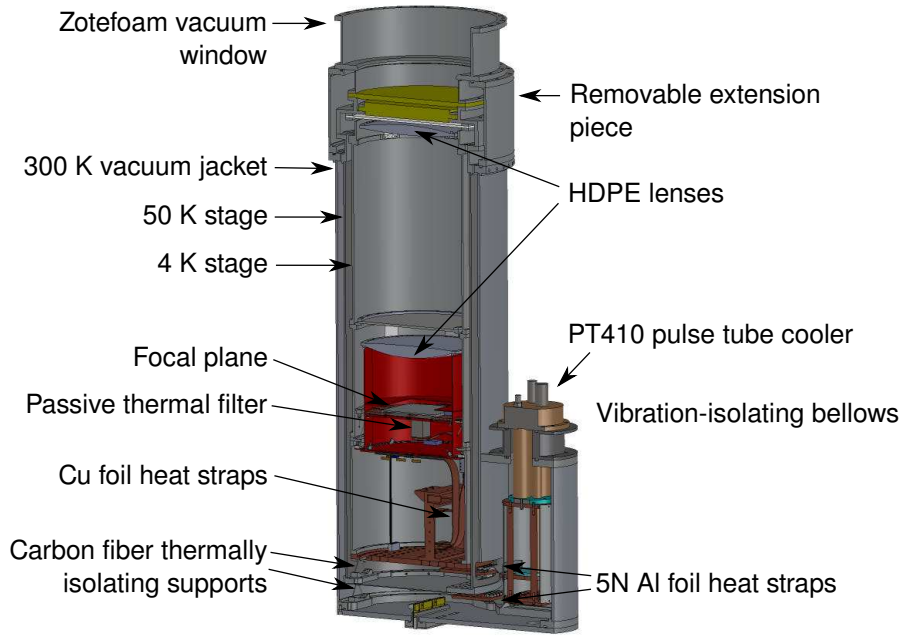}
   \end{tabular}
   \end{center}
   \caption[example]
   { \label{fig:keckdewar}
\keck\ cryostat design.  Major changes relative to \bicep2 are the addition of the
extension piece; the removal of the liquid helium reservoir and vapor cooled shield,
with correspondingly narrower 300~K vacuum jacket; and addition of the PT410 pulse
tube enclosure, with vibration isolation and heat straps.
}
   \end{figure}

\bicep2 is a small-aperture refracting telescope that observes the CMB at $150~\mathrm{GHz}$ from the South Pole.
In many respects it follows the pattern of \bicep, which was designed to achieve high sensitivity with very low
systematics on degree angular scales.  The telescope contains relatively simple optics:
two, on-axis lenses made
from high-density polyethylene, with an aperture of $26.4~\mathrm{cm}$ and beams of 0.5$^\circ$ FHWM.
This arrangement allows cooling of the optical components to 4~K for stability and low loading;
efficient transportation and
loading; and far-field ($>50~\mathrm{m}$) beam characterization using microwave sources raised on masts.
The \keck\ expands this design in a modular way, by placing several \bicep2-style telescopes in a single
mount.  Each telescope observes in a single frequency band, with optics optimized for this frequency.
\bicep2 and the current \keck\ receivers all observe at $150~\mathrm{GHz}$, but future \keck\ receivers
may observe at $100$ and $220~\mathrm{GHz}$.

\bicep2 and the \keck\ observe the same primary CMB field as \bicep, an 800 deg$^2$ patch of sky in the
so-called Southern Hole.  This area has unusually low dust and synchrotron emission,
and is expected to have Galactic foreground confusion equivalent to $r\lsim 0.03$.

\bicep2 has pioneered a new detector technology in order to increase detector count and
instrument sensitivity.
The Caltech/JPL focal planes in \bicep2 and the \keck\ combine the beam-defining slot-antenna networks,
band-defining filters, absorber, and transition-edge sensor (TES) detectors on the same
silicon detector tiles.  The actual \bicep2 focal plane and the far-field beam pattern of a single
\keck\ receiver are shown in Fig.~\ref{fig:b2fpubeams}.
This detector technology allows scaling to higher detector count by
replacing hand-assembly of components in preceding experiments with monolithic fabrication.
The Ti and Al TES
elements are suspended on a SiN membrane along with a gold resistive absorber.  The Ti transition temperature
$T_c$ is in the range 470-530~mK.  In order to facilitate beam characterization under higher loading,
the Ti sensors are in series with an Al TES with $T_c$ of 1.2~K.  Each telescope contains four detector
tiles, each with 64 dual-polarization elements in an $8\times8$ array.  Two of the detector
pairs on each tile are dark sensors with no coupling to antennas, for a total of 496 optically coupled,
$150~\mathrm{GHz}$ detectors and 16 dark TES detectors per receiver.\footnote{The detector counts for
future $100$ and $220~\mathrm{GHz}$ receivers will be different.  \bicep2 has two fewer dark pairs,
for a total of 12 dark and 500 optically coupled TESs.}
The detectors are designed to be read out with NIST three-stage SQUID hardware and the University of
British Columbia Multi-Channel Electronics (MCE) system, with time-domain multiplexing.
The technology of
antenna-coupled TES detectors with multiplexed readout has allowed \bicep2 to increase the detector count of \bicep\ by a
factor of five.  The \keck, with five telescopes, increases this detector count by a further factor of five.

Each telescope is housed within a cryostat that provides cooling to 4~K.  The \bicep2 and \keck\ designs have
progressively eliminated consumable cryogens, for which the Amundsen-Scott South Pole Station has limited
storage and transportation capacity.  \bicep2 sits inside a liquid helium cryostat from
Redstone Aerospace\footnote{http://www.redstoneaerospace.com/}, which is similar to the \bicep\ dewar but
replaces a liquid nitrogen jacket with a vapor-cooled shield.  The helium reservoir has a capacity of 100~L,
and consumes about 22~L/day during routine observing.  The \keck\ cryostats (Fig.~\ref{fig:keckdewar}), manufactured by
Atlas UHV\footnote{http://www.atlasuhv.com/}, have several major changes relative to \bicep2.
They add a removable extension piece for a half-wave plate, eliminate the liquid helium reservoirs, and add a housing
for a Cryomech PT410 two-stage pulse tube cooler mounted on vibration-isolating bellows.
The pulse tube coolers are rated to
maintain $T<50~\mathrm{K}$ on the first
stage, with loading up to 45~W, and $T<3.5~\mathrm{K}$ on the second stage, with loading of $0.5~\mathrm{W}$.
The thermal linkage from the pulse tube cold head to the 4~K plate of the telescope is made from
a stack of high-purity Al foil, which provides excellent thermal conduction while isolating the telescope from
vibration of the pulse tube.
The \keck\ cryocoolers and thermal linkage have demonstrated excellent performance after deployment.
During standard observing, the cooler achieves temperatures
$\leq3.0~\mathrm{K}$ at the pulse tube second stage and $\leq3.6~\mathrm{K}$ on the telescope side of the Al strap.
The first stage of the cryocooler maintains $\sim40~\mathrm{K}$.

Cooling below $4~\mathrm{K}$ is provided by a Duband three-stage $^4$He/$^3$He/$^3$He sorption refrigerator
operating in a closed cycle.  \bicep2 has made several improvements in the thermal architecture in order to
improve stability and reduce microphonic pickup.  These have also been adopted by the \keck.
The focal plane is cooled to base temperature through copper-foil heat straps and
a passive thermal filter and copper heat straps that
link it to the lowest-temperature stage of the fridge.  The passive thermal filter is a stainless steel block,
$5.5~\mathrm{cm}\times2.5~\mathrm{cm}\times2.5~\mathrm{cm}$.  The focal plane is supported by a truss structure
of carbon-fiber rods, providing good thermal isolation from the $4~\mathrm{K}$ stage.\cite{runyan08}
Temperature control is achieved using resistive heaters and thermometers on the focal plane and at the
fridge side of the passive thermal filter, servoed to suppress thermal fluctuations and to keep the focal plane
at a temperature near 280~mK.
The thermal architecture of \bicep2 has performed well.  Estimates of the level at which thermal fluctuations
on the focal plane enter CMB maps are lower than the corresponding estimates for \bicep, and the overall
thermal stability while scanning has met our design goals.

The \bicep2 refrigerator can be recycled in slightly under 4 hours, after which it remains at a stable base temperature for
at least 80 hours.  The recycling operation and liquid helium fill are performed every three sidereal days, during a cryogenic
service block of six hours.
Several times per month, during clear weather, the cryogenic service block is also used for star observations,
which are analyzed to characterize the telescope pointing and particularly to detect any shifts in the pointing.
Star observations are performed several times per month, during 
In the \keck\ performance varies by receiver, and the observing schedule must accommodate the
receiver with the longest service time and shortest hold time.  The fridge recycling requires up to
six hours, generally completing by the end of the six-hour service block.  For the \keck\ this cryogenic service
block occurs every two sidereal days.

\section{SENSITIVITY IMPROVEMENTS}
Several upgrades since the initial deployment of \bicep2 and the \keck\ have substantially increased
their sensitivities.  Although some details differ between the two experiments, these improvements have
proceeded along parallel paths.  We have reduced aliased noise by increasing the multiplexing rate,
and optimized the detector biases for sensitivity and to minimize the number of unresponsive detectors.

\bicep2 has continued to run since deployment in a single cooldown without warming up or breaking vacuum.
The \keck\ underwent major upgrades during the 2011-12 austral summer.  This included the
addition of two new receivers, the replacement of the focal plane and detectors in one receiver,
and the removal of cold half-wave plates from two
of the receivers.

\subsection{Detector bias improvements}
In both experiments, the TES detector bias voltages were chosen conservatively at initial
deployment.  These were revisited after several months of observation,
with the goals of improving
overall sensitivity and eliminating instabilities in some detectors.
The new, optimized biases were chosen after study of the
detector responsivity and stability in situ under good, winter observing conditions.

The detector bias is provided by the Multi-Channel Electronics (MCE)
system.  The TESs are biased with a common current bias for the
32 detectors\footnote{Each column also contains one open-input SQUID, for 33 total channels.}
in each multiplexing column.  This current bias is converted into a voltage
bias by a shunt resistor in parallel with the TES.
This scheme allows some optimization of the biases for variation between
detectors, since each multiplexing column can have a different bias voltage.
The bias settings for each column must be chosen to give good responsivity and stability for all the detectors
sharing that bias current.
The responsivity is highest when the detector is within the middle portion of the superconducting-normal
transition, i.e. the fractional resistance $R/R_{normal}$ is not too close to zero or one.  When the
resistance is too low, the detector may also enter a state of unstable electrothermal
feedback, which renders it unusable and can cause crosstalk signals on nearby channels.
The optimal bias value depends on detector parameters such as thermal conductance and superconducting
transition temperature.  These are relatively consistent among the detectors in a single multiplexing column,
which are always located on the same physical detector tile.
The optimal bias values also depend on atmospheric loading.  Under low-loading (winter) conditions, the highest
responsivity is attained at relatively high bias voltage.  Under high-loading (summer) conditions, on the
other hand, lower bias voltages are required to keep $R/R_{normal}$ well below one.

At deployment of \bicep2, the TES biases were chosen conservatively, during the austral summer.
They were deliberately somewhat lower than the
optimal per-column values, in order to avoid regions of decreasing responsivity.
\cite{brevik10}  During mid-2010, additional test data
were taken at a variety of biases under winter conditions of low atmospheric loading.  These studies led to
improved bias selections, with an expected improvement of 10-20\% in mapping speed.
The new biases are the result of a careful balance between achieving the best
sensitivity for as many detectors as possible, while minimizing the number of detectors that are either
unresponsive or unstable.  This optimization is done in the total NET of each column, rather than in a
per-detector mean.  Therefore, it has sometimes been acceptable to allow detectors with unusual bias
requirements to fall off the transition entirely, in exchange for achieving better sensitivity in the
others.
The SQUID feedback is switched off for some unstable detectors, so that their
noisy feedback signal does not affect
neighboring channels through the known crosstalk mechanisms of the readout system.\cite{dekorte03}

The new detector biases were adopted on September 14, 2010, and have been used through the remainder of the
2010 season, continuing through the full 2011 season and the 2012 season now in progress.  The \keck\ has followed
a similar procedure, with conservative bias settings at deployment replaced by more nearly optimal
values after study under winter loading.\cite{kernasovskiy12}

\subsection{Faster multiplexing}

The multiplexing configuration used by \bicep2 in 2010 has been described in detail in a previous
SPIE conference.\cite{ogburn10}
The row visit rate was 15~kHz, so any noise above 7.5~kHz that was not adequately suppressed
before digitization could be aliased and impact the low-frequency region containing the CMB signal.
Accordingly, the readout includes $1.35~\mu\mathrm{H}$ bandwidth-limiting inductors on the Nyquist
chips.  The resulting LR circuit is a single-pole filter with cutoff
at $5$--$6\mathrm{kHz}$.\footnote{The $R$ of the LR circuit is the resistance of the TES itself, so
the exact Nyquist frequency depends on the detector bias.}
Because the pole of this slow-rolloff filter is not far below the Nyquist filter of the
multiplexing system in 15-kHz configuration, some amount of noise aliasing will occur.

\begin{table}[t]
\caption{Multiplexing parameters used by \bicep2 and the \keck}
\label{tab:mux}
\begin{center}
\begin{tabular}{|l|l|l|} 
\hline
\rule[-1ex]{0pt}{3.5ex}  & 15~kHz & 25~kHz \\
\hline
\rule[-1ex]{0pt}{3.5ex}  Raw ADC sample rate & 50~MHz  & 50~MHz \\
\hline
\rule[-1ex]{0pt}{3.5ex}  Row dwell & 98~samples & 60~samples  \\
\hline
\rule[-1ex]{0pt}{3.5ex}  Row switching rate & 510~kHz & 833~kHz \\
\hline
\rule[-1ex]{0pt}{3.5ex}  Number of rows & 33 & 33 \\
\hline
\rule[-1ex]{0pt}{3.5ex}  Row revisit rate & 15.46~kHz & 25.25~kHz \\
\hline
\rule[-1ex]{0pt}{3.5ex}  Internal downsample & 150 & 140 \\
\hline
\rule[-1ex]{0pt}{3.5ex}  Output data rate per channel & 103~Hz & 180~Hz \\
\hline
\rule[-1ex]{0pt}{3.5ex}  Software downsample & 5 & 9 \\
\hline
\rule[-1ex]{0pt}{3.5ex}  Archived data rate & 20.6~Hz & 20.0~Hz \\
\hline
\end{tabular}
\end{center}
\end{table}


A detailed noise model and comparison with observed noise levels were presented in an earlier
SPIE conference.\cite{brevik10}  It was found that, under 4--6~pW loading, the total noise-equivalent power (NEP)
 at low frequency was
$50$--$60~\mathrm{aW}/\sqrt{\mathrm{Hz}}$.  Photon noise was the largest contribution, with
phonon noise (thermal fluctuation noise) at a comparable but lower level.  Excess noise was
also seen, and became dominant above 100 Hz.  When multiplexing at 15~kHz, this excess noise was
partly aliased into the low-frequency ($<5~\mathrm{Hz}$) science band, raising the total
NEP to $70$--$80~\mathrm{aW}/\sqrt{\mathrm{Hz}}$.

Although this noise aliasing was understood at the time of deployment, it was not possible to shift the
Nyquist filter because higher-inductance Nyquist chips were not available.
Instead, it was anticipated that the noise aliasing could be ameliorated by shifting to a faster multiplexing
rate.  This requires a shorter dwell time on each row.  The dwell time must be long enough that
all transients associated with row switching have fully settled before moving on to the next row.
These transients are a form of crosstalk between detectors in neighboring multiplexing rows.\cite{dekorte03}
\bicep2 continued to operate at 15-kHz readout until it could be demonstrated that a faster
multiplexing rate would not increase row-switching crosstalk to unacceptable levels.

These studies of crosstalk and multiplexing rate were performed in late 2010, and it was found that
(after making other adjustments to the SQUID bias settings)
the readout rate could be increased without significant increase in crosstalk.
As part of the 2010--11 summer upgrades, \bicep2 began to run with 25~kHz row switching.  The
new multiplexing parameters are shown in Table~\ref{tab:mux}.  
We expected the reduction in aliased noise to yield a sensitivity gain of about 20\% in the
\bicep2 2011 and 2012 data.\cite{brevik10}  The actual gains in sensitivity are discussed in
Sec.~\ref{subsec:sens}.

The \keck\ was able to obtain Nyquist chips with $2.0~\mu\mathrm{H}$
inductors for four of its five receivers (all except rx0).  This reduced the noise aliasing,
but there remains a further benefit from faster readout.  The initial deployment
continued to use the well-tested 15-kHz multiplexing configuration, pending successful outcome of the
25-kHz change in \bicep2 and study of crosstalk in the slightly different \keck\ readout hardware.
Since the faster readout was found also to work well in \keck\, with acceptable levels of crosstalk,
the \keck\ row-switching rate was increased to 25~kHz during the 2011-12 summer.  Although the
\keck\ uses newer versions of the MCE hardware and firmware, allowing greater
freedom in the choice of multiplexing parameters, we have adopted the same values used in \bicep2 for
consistency and because they have already been demonstrated to work well.  As a result, the
\bicep2 2010 and \keck\ 2011 data sets use the same 15-kHz readout settings, and the \bicep2 2011-12 and
\keck\ 2012 data sets use the same 25-kHz readout settings.

\subsection{Temperature control}
The \bicep2 and \keck\ focal planes are cooled to
$250~\mathrm{mK}$ by Duband three-stage $^4$He/$^3$He/$^3$He
sorption refrigerators operating in a closed cycle.  
The thermal path between the refrigerator and the focal plane
has been designed to isolate the focal plane
assembly from microphonic noise and
from short-period thermal fluctuations.  The $4$-K, $350$-mK, and
base temperature stages of the refrigerator are connected to the truss
assembly behind the focal plane by flexible,
straps of layered copper foil.  The $250$-mK strap is attached to a
stainless steel passive thermal filter that effectively isolates the 
focal plane from thermal fluctuations faster than about
$1300~\mathrm{s}$.\cite{ogburn10}

The temperatures of the thermal straps, focal plane, and detector tile are
monitored with neutron-transmutation doped (NTD) germanium thermistors.  These are paired with heaters that provide
for active thermal control in several possible configurations.  In standard
observing, the primary thermal control is on the fridge side (``dirty'' side)
of the passive thermal filter.  This control loop is tuned to have $\lsim1$-Hz
bandwidth in order to remove temperature variations produced in the refrigerator
as the telescope scans.  The remaining small fluctuations are strongly attenuated
by the stainless steel filter.  The heater on the focal plane copper plate is
also servoed.  This ``slow'' control loop has a time constant of $\sim30~\mathrm{s}$ and is
intended to keep the detector tiles at the designated operating temperature,
rather than to suppress thermal fluctuations.

The effectiveness of this thermal architecture is assessed using NTD germanium thermometers
mounted on the detector tiles themselves.  In the first year of \bicep2 observing,
the thermometry suffered from elevated noise due to microphonic pickup in
the high-impedance wiring and excess noise in the room-temperature readout
electronics.  This additional noise had minimal effect
on the thermal stability of the detector tiles, as it was well above the
bandwidth of the thermal control loop on the focal plane.  However, it did
impede efforts to characterize the thermal performance of the system.

For the second and third seasons of \bicep2, the alternating current thermistor bias
signals have been increased in frequency from $50~\mathrm{Hz}$ to $100~\mathrm{Hz}$
in order to better avoid microphonic noise.  An upgrade to the readout firmware allowed
the use of nulling resistors and increased bias amplitudes (typical values $15~\mathrm{mV}$ in 2010
to $40~\mathrm{mV}$ in 2011--12) for higher signal-to-noise.
The \keck\ uses a later hardware version of the digital electronics,
which eliminates much of the excess noise seen by \bicep2. 


\subsection{Additional \keck\ receivers}

The \keck\ initially deployed three receivers for the 2011 season, known as rx0, rx1, and rx2.
Two new receivers
were added during the 2011--12 summer, rx3 and rx4.  In
addition, the detectors in one receiver (rx1) were found to have degraded
performance relative to the other recievers.  Its focal plane unit was replaced as part
of the year-two upgrades.  Although the 2011 and 2012 receivers have all contained
detectors at $150~\mathrm{GHz}$, the modular \keck\ design also allows the possibility of
swapping in receivers at $100~\mathrm{GHz}$ and $220~\mathrm{GHz}$ in the future.

\subsection{Half-wave plates}
The \keck\ cryostat design included a removable extension to allow for additional thermal
filters or for a stepped half-wave plate.  During the 2011 season, two of the receivers,
(rx1 and rx2) were actually equipped with sapphire half-wave plates cooled to $6~\mathrm{K}$,
and wave-plate rotations were incorporated into
an observing strategy based on the \bicep2 pattern.\cite{ogburn10}
Before each sidereal day's observations, the telescope was rotated about the boresight axis,
so that each
day covered one of the four boresight orientations 68$^\circ$,
113$^\circ$, 248$^\circ$, 293$^\circ$.  (These are identical
to the boresight angles used by \bicep2, and allow separation of the Q and U Stokes
parameters as well as providing a 180$^\circ$ jackknife for characterization and
control of systematics).
After four days of data-taking, covering all four boresight angles, the half-wave plates were
stepped by 45$^\circ$.  This procedure is repeated to cover the four half-wave-plate
orientations 0, 45$^\circ$, 90$^\circ$, 135$^\circ$.

This observing pattern was intended to address the known differential pointing common to
the antenna-coupled detectors ($0.1\pm0.04^\circ$ in \bicep2\cite{aikin10},
typically $0.05^\circ$ or smaller in the \keck\cite{vieregg12}).  The wave-plate rotation periodically exchanges the roles
of the two detectors in each pair, while the boresight orientation and all other aspects
of the observation remain fixed.  Jackknife combinations of maps made at different
half-wave plate orientations could be used to better characterize the differential pointing,
and coadding maps over half-wave plate orientations would suppress the temperature-to-polarization
leakage caused by differential pointing.

In the 2011 data set, the two receivers with half-wave plates were 
found to suffer from degraded optical responsivity and additional beam 
systematics that were difficult to understand.  These issues disappeared 
when the half-wave plates were removed for 2012 observing.

\section{PERFORMANCE AND RESULTS} 

\bicep2 has now completed two years of observing and is at the midpoint of its third year.
We present measures of the observing efficiency and sensitivity, particularly showing the
significant gain in sensitivity resulting from the upgrades described above.  We also 
present preliminary maps from \bicep2, including maps of the Stokes Q and U
parameters and $E$ and $B$ polarization.

The \keck\ has completed one year of observing and is now in the middle of its second year,
with greatly improved sensitivity thanks to the additional receivers and other upgrades.
A detailed discussion of the \keck\ 2012 sensitivity, along with preliminary maps, may
be found in a companion paper (Kernasovskiy \etal, this volume\cite{kernasovskiy12}).

   \begin{figure}
   \begin{center}
   \begin{tabular}{c}
   \includegraphics[height=5cm]{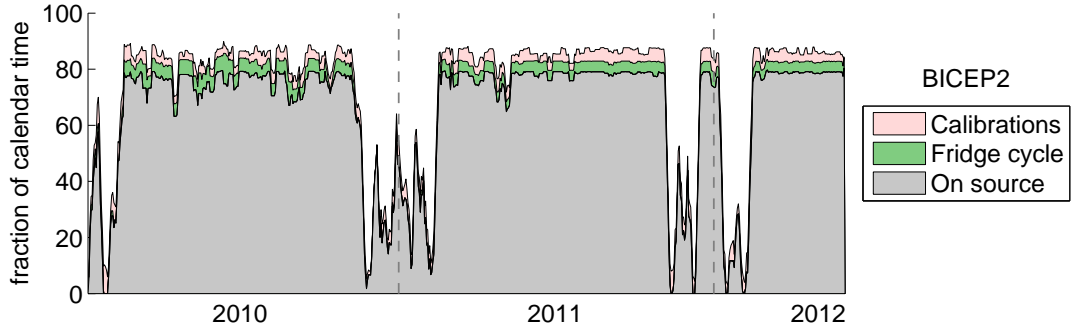}
   \end{tabular}
   \end{center}
   \caption[example]
   { \label{fig:livetime}
Time spent by \bicep2 in CMB scans, regular calibrations, and refrigerator cycling since January 1, 2010.
During austral summers (November-February), observing schedules have been interspersed with beam mapping
and other tests and calibrations.  During the austral winter, on-source efficiency has been high,
never falling far below the ideal 79.2\% in the late 2011 and early 2012 observing periods.
The remaining time not categorized in this figure includes scheduled pauses for detector biasing,
thermal settling, and telescope slewing; spare time in the 6-hour cryogenic block;
and time spent in summer calibration and testing tasks.
}
   \end{figure}

\subsection{Observing efficiency}
\bicep2 has a three-day cryogenic hold time and an ideal on-source fraction of 79.2\%.
This is the fraction of the time that the telescope would spend on the field if it continuously
executed the standard three-day observing schedule, without interruption.
(Because the \keck\ has
a two-day observing cycle, its maximum observing efficiency is slightly lower, 70.0\%.)  Observing during the austral
winter has generally been very close to this ideal value, as shown in Fig.~\ref{fig:livetime}.
In particular, the on-source fraction since mid-2011 has never fallen far below the ideal value except during
summer beam-mapping and other calibrations.
The on-source fraction includes observations of a galactic field, which is allotted seven hours out of each
three-day \bicep2 observing schedule.
In all, \bicep2 has integrated on the primary CMB field for
4395 hours in 2010\footnote{from February 1}, 4891 hours in 2011\footnote{from January 1}, and 2034 hours so far
in 2012\footnote{January 1-May 31}.  These on-source figures are before the application of weather and
data-quality cuts, and include the scan turn-arounds, which are not used for mapmaking and account for
22.3\% of on-source time.

\subsection{Sensitivity}
\label{subsec:sens}
The sensitivity of \bicep2 has been calculated for three
time periods: early 2010, late 2010, and 2011-12.  The early 2010 period is as deployed, with
$15$-kHz multiplexing and conservative TES biases.  The late 2010 period still has $15$-kHz
readout, but uses the improved TES biases for higher signal-to-noise ratio.  The 2011-12 period
uses the faster $25$-kHz readout with lower noise, and the improved detector biases.

Sensitivity has been expressed as a noise-equivalent temperature (NET) as determined in two
ways: from time-stream noise in the 0.1--1 Hz signal band; and from the variance of pixel values
in maps differenced between left-going and right-going scans (a scan-direction jackknife map).
The methodology for calculating the NETs is as in Kernasovskiy \etal\cite{kernasovskiy12} in
these proceedings.
The results are broadly consistent.  The per-detector NETs are given in Table~\ref{tab:netperdet},
and the overall instrument NETs in Table~\ref{tab:netinst}, after Brevik \etal\cite{brevik11,brevikthesis}.
Note that the NET in the late-2010 period has been calculated only from timestream noise spectra,
and not from maps.

\begin{table}[t]
\caption{Per-detector \bicep2 NET}
\label{tab:netperdet}
\begin{center}
\begin{tabular}{|l|l|l|} 
\hline
\rule[-1ex]{0pt}{3.5ex}  Period & Time stream NET, $\mu\mathrm{K}\sqrt{\mathrm{s}}$ & Map NET, $\mu\mathrm{K}\sqrt{\mathrm{s}}$ \\
\hline
\rule[-1ex]{0pt}{3.5ex}  Early 2010 & 433  & 422 \\
\hline
\rule[-1ex]{0pt}{3.5ex}  Late 2010 & 379 & - \\
\hline
\rule[-1ex]{0pt}{3.5ex}  2011-12 & 316 & 313  \\
\hline
\end{tabular}
\end{center}
\end{table}

\begin{table}[t]
\caption{\bicep2 instrument NET}
\label{tab:netinst}
\begin{center}
\begin{tabular}{|l|l|l|} 
\hline
\rule[-1ex]{0pt}{3.5ex}  Period & Time stream NET, $\mu\mathrm{K}\sqrt{s}$ & Map NET, $\mu\mathrm{K}\sqrt{s}$ \\
\hline
\rule[-1ex]{0pt}{3.5ex}  Early 2010 & 22.1  & 21.5 \\
\hline
\rule[-1ex]{0pt}{3.5ex}  Late 2010 & 19.1 & - \\
\hline
\rule[-1ex]{0pt}{3.5ex}  2011-12 & 15.9 & 15.8  \\
\hline
\end{tabular}
\end{center}
\end{table}

\subsection{Polarization maps}
\begin{figure}
   \begin{center}
   \subfigure[]{
     \includegraphics[height=6cm]{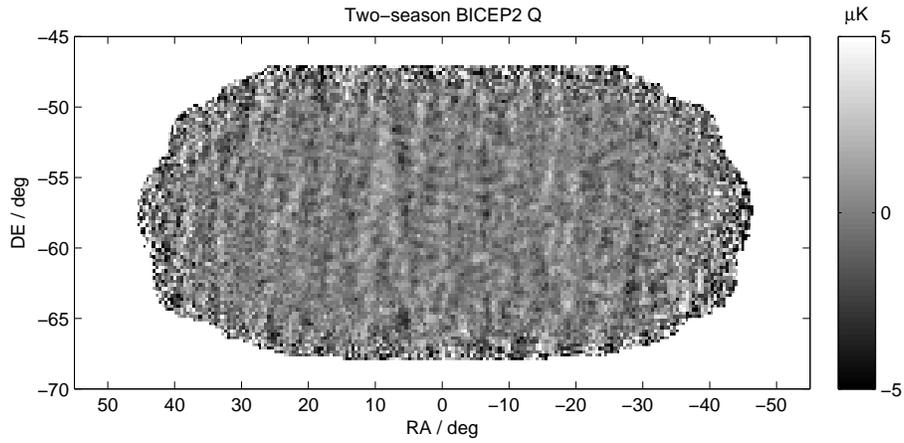}
     \label{fig:qmap}
   }
   \vspace{0mm}
   \subfigure[]{
     \includegraphics[height=6cm]{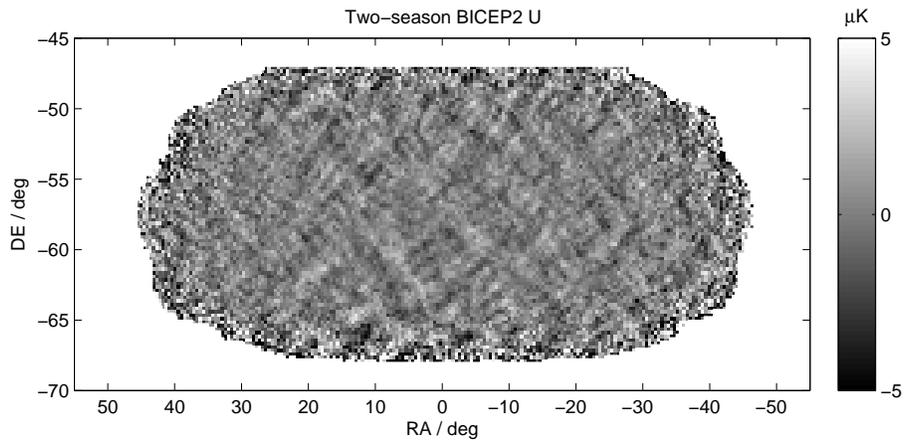}
     \label{fig:umap}
   }
   \caption{\bicep2 polarization maps of the Stokes Q parameter (a) and Stokes U parameter (b) from full-season 2010 and 2011 data.
            Map pixels are 0.25$^\circ$ in declination and 0.47$^\circ$ in right ascension.
            The maps have not been apodized or smoothed in $\ell$.  The structure of the CMB polarization is immediately apparent by
            eye: power is predominantly vertical and horizontal in Q, and diagonal in U.  This pattern corresponds to $E$-mode 
            polarization with very little power in $B$-modes.
            }
   \end{center}
   \end{figure}

In Fig.~\ref{fig:qmap} and \ref{fig:umap} we show \bicep2 polarization maps made
with the full-season 2010 and 2011 data sets.  The data analysis is
similar to that used in \bicep\cite{chiang10}.  These maps do not include
analysis techniques being developed\cite{deproject} to remove the leaked polarization signal caused by
differential pointing of the two detectors in a polarization-pair.  The $E$-mode
signal around $\ell=400$ is readily apparent as a set of horizontal and vertical
structures in the Stokes-Q map and diagonal structures in the Stokes-U map.  The prevalence of $E$-mode
power at this angular scale results from the intrinsic shape of the $E$-mode CMB spectrum (which
increases with $\ell$ below $ell=1000$) and the \bicep2 beams, which are not sensitive to power at higher $\ell$.
We have estimated the map noise level by making separate maps from the left-going and
right-going scans, differencing them, and dividing by two.  This scan-direction jackknife map should contain
noise at the same level as the fully coadded map, but no signal.
From the pixel variance in the well-covered central region, we calculate noise levels
in the Stokes Q and U maps of 0.139 and 0.138~$\mu$K in square-degree pixels
(8.3~$\mu$K $\cdot$ arcmin).

\begin{figure}
    \begin{center}
    \subfigure[]{
      \includegraphics[height=6cm]{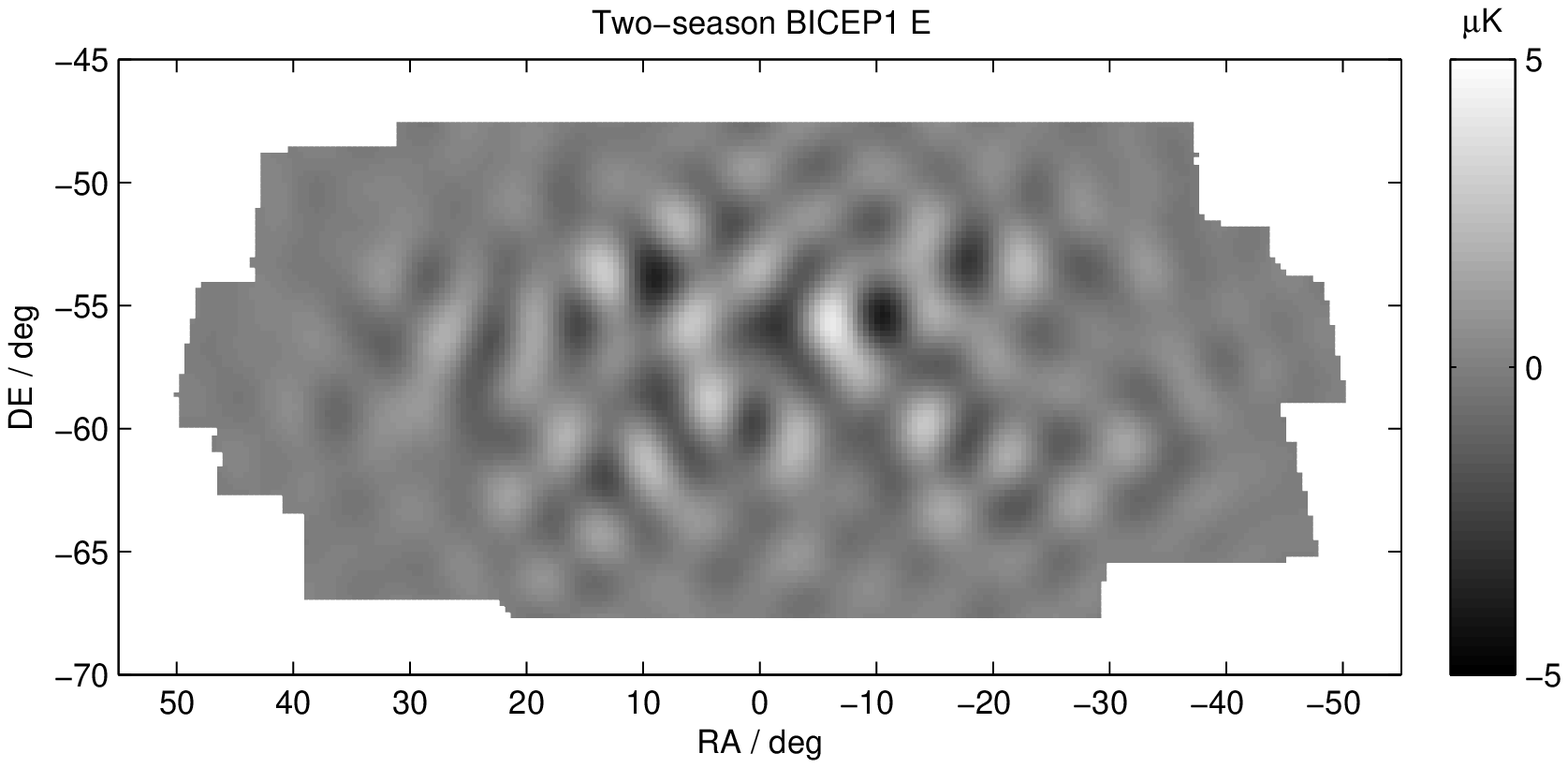}
      \label{fig:b1e}
    }
    \vspace{0mm}
    \subfigure[]{
      \includegraphics[height=6cm]{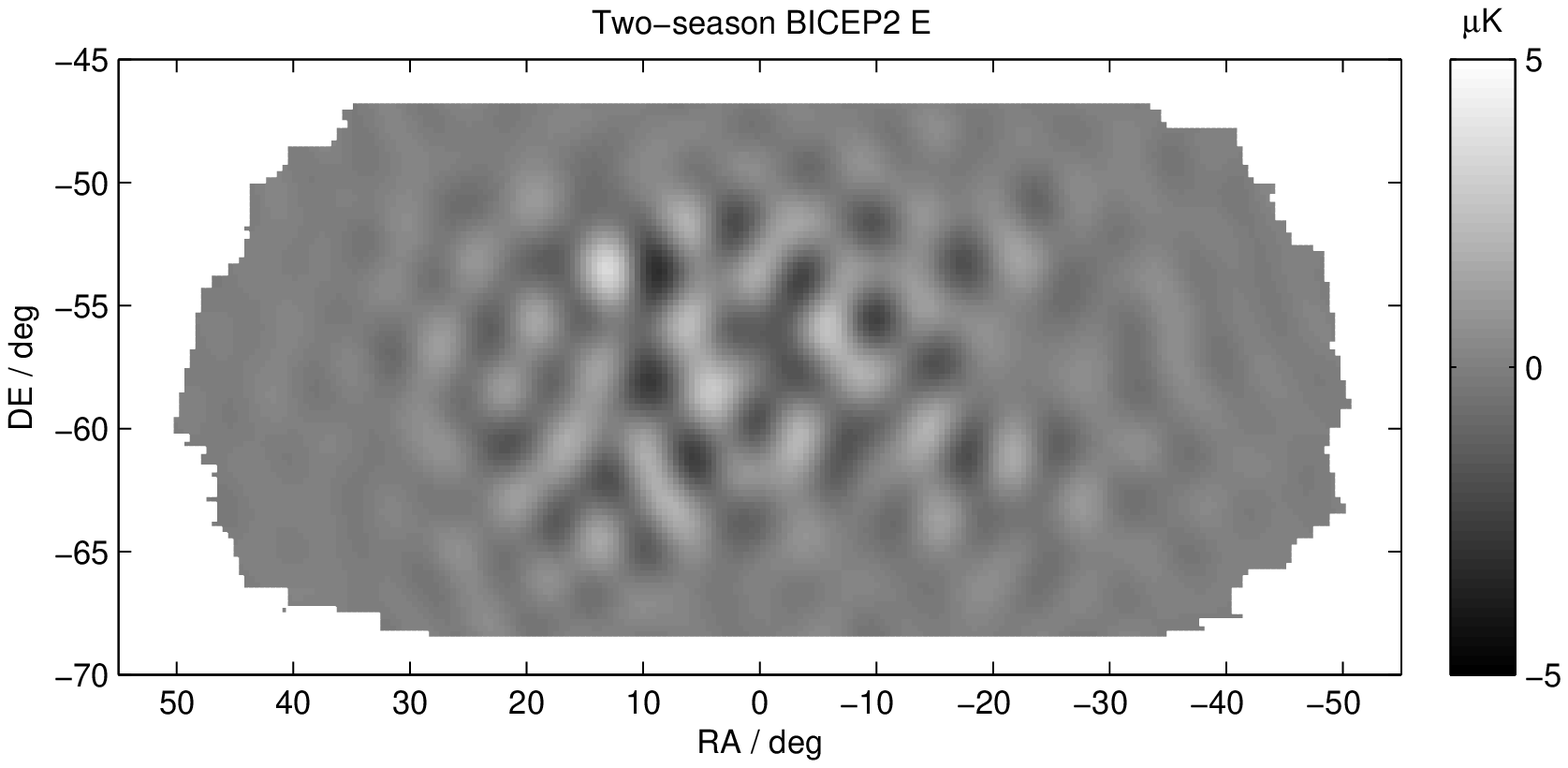}
      \label{fig:b2e}
    }
    \caption{Maps in $E$-mode polarization made from two full seasons of (a) \bicep\ data
             (2006--07) and (b) \bicep2 data (2010--11).
             These maps are apodized by the pixel variance map and include
             modes in the range $50\leq\ell\leq120$.
               The $E$-mode maps from both \bicep\ and
             \bicep2 are dominated by true CMB polarization, and hence show common features.\label{fig:b12e}}
    \end{center}
    \end{figure}

\begin{figure}
    \begin{center}
    \subfigure[]{
      \includegraphics[height=6cm]{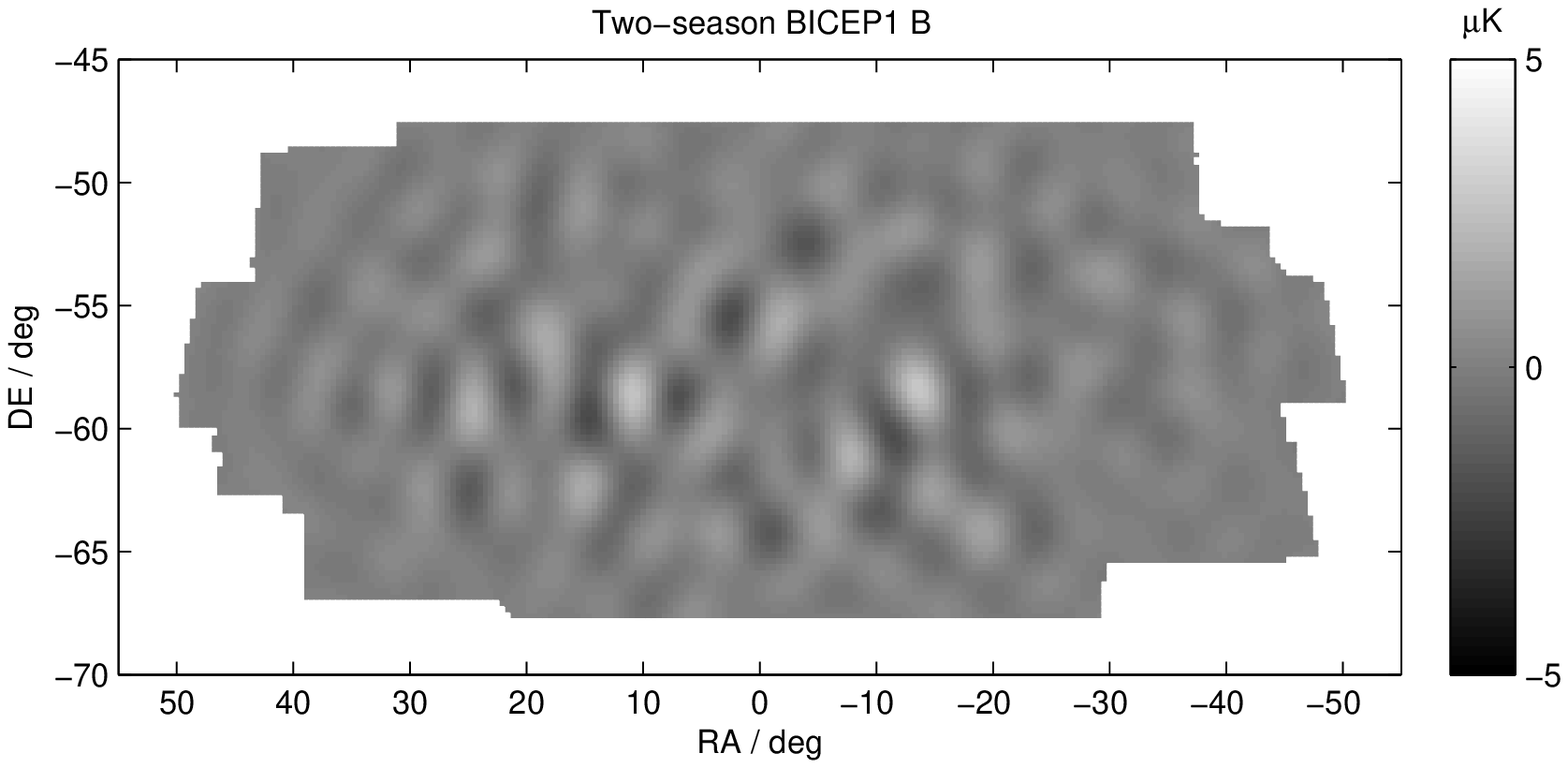}
      \label{fig:b1b}
    }
    \vspace{0mm}
    \subfigure[]{
      \includegraphics[height=6cm]{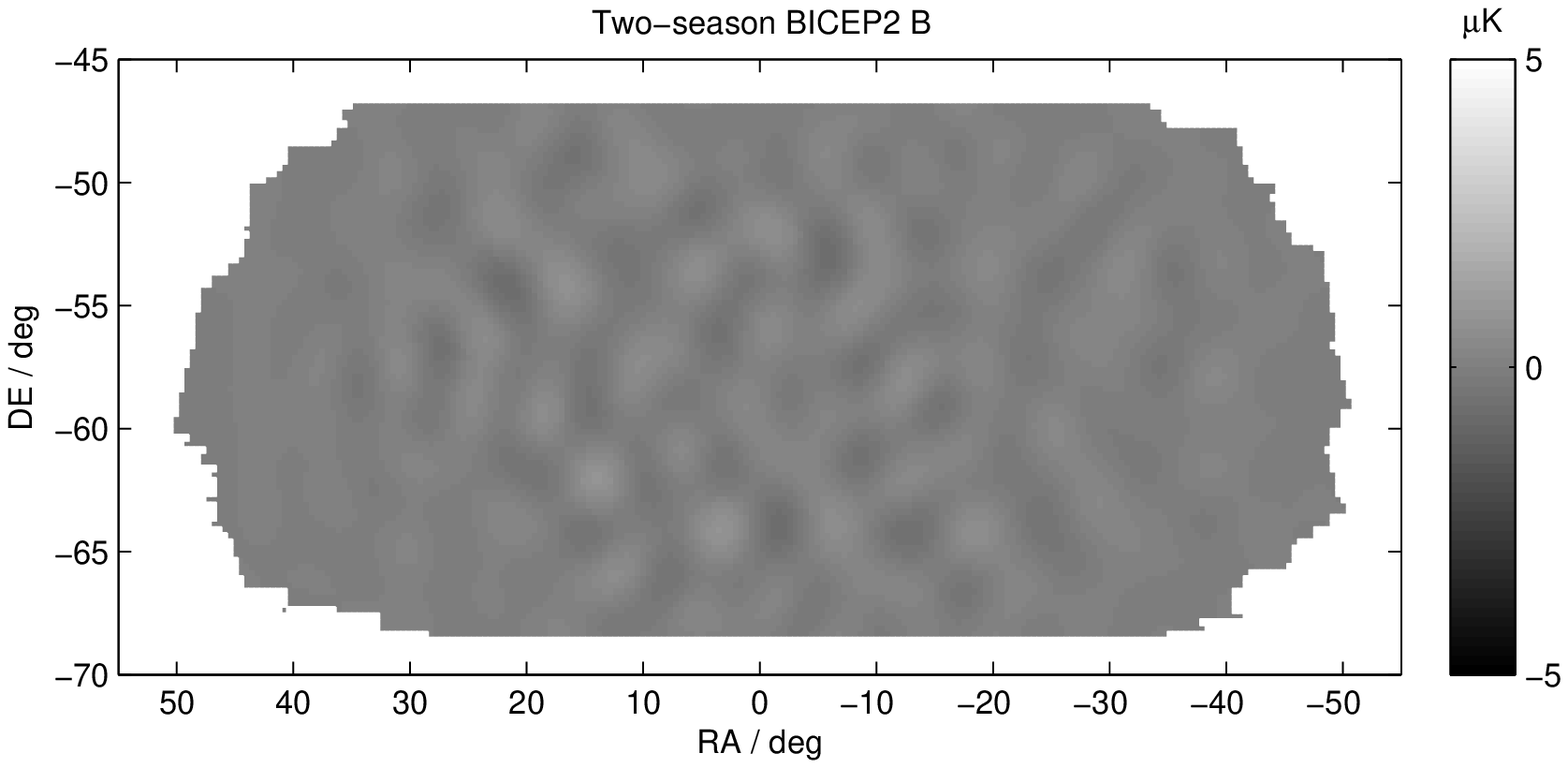}
      \label{fig:b2b}
    }
    \caption{Maps in $B$-mode polarization made from two full seasons of (a) \bicep\ data
             (2006--07) and (b) \bicep2 data (2010--11).
             These maps are apodized by the pixel variance map and include
             modes in the range $50\leq\ell\leq120$.
             The \bicep\ $B$-mode map is dominated by noise.  The greater sensitivity and
             mapping speed of \bicep2 result in a $B$-mode map with much smaller amplitude.
             \label{fig:b12b}}
    \end{center}
    \end{figure}

Fig.~\ref{fig:b12e} and \ref{fig:b12b} shows the $E$-mode and $B$-mode
patterns respectively reconstructed from two seasons of \bicep\ data\cite{chiang10}
and from the two-season Stokes Q and U maps shown in Fig.~\ref{fig:qmap}
and \ref{fig:umap}.  These have been apodized using the variance per map pixel,
and filtered to the angular scale $50\leq\ell\leq120$.
The $E$-mode maps are similar, and the lower level of power
in the $B$-mode map shows the improved sensitivity of \bicep2 relative to \bicep.


\section{CONCLUSION} 
\bicep2 has completed two years of observation at the South Pole, and is currently in the middle of its
third season.  The \keck\ has completed its first season, with three receivers, and is now in the middle
of its second season, with five receivers.  Both have made substantial improvements to their sensitivity,
through optimization of the detector biases and readout rate, and through the addition of new \keck\
receivers and replacement of detectors.  \bicep2 has accumulated a total of 11320 hours on the primary
CMB field.
We have presented for the first time preliminary \bicep2 maps of the primary CMB field in polarization, showing the
improvement in sensitivity over the original \bicep\ experiment.

\acknowledgments     
The \bicep2 and \keck\ projects have been made possible through
support from the National Science Foundation (grant Nos. ANT-1044978 /
ANT-1110087), the Keck Foundation, the Canada Foundation for Innovation,
and the British Columbia Development Fund. 
Detector development has been made possible by the generous support of the Gordon and Betty Moore Foundation.
RWO gratefully acknowledges support from the Kavli Institute for Particle Astrophysics and Cosmology.
We are grateful to Steffen Richter as our 2010-12 \bicep2 winter-over, and to Robert Schwarz
as our 2011-12 \keck\ winter-over.
The \bicep2 and \keck\ teams would also like to thank the South Pole Station staff for logistical support.
We thank our \bicep\ and \spider\ colleagues for useful discussions and shared expertise.


\bibliography{report}   
\bibliographystyle{spiebib}   

\end{document}